\documentclass[12pt]{article}
\usepackage{epsfig}

\parskip=\medskipamount	
\renewcommand{\thesection}{\arabic{section}}

\catcode`@=11

\@addtoreset{equation}{section}
\@addtoreset{equation}{subsection}
\def\theequation{\ifnum\value{section}=0 \arabic{equation}\ignorespaces
\else \ifnum\value{section}=-1 A.\arabic{equation}\ignorespaces
\else \ifnum\value{subsection}=0 \thesection.\arabic{equation}\ignorespaces
\else \thesection.\arabic{subsection}.\arabic{equation}\ignorespaces
                             \fi
                        \fi
                   \fi}

{\catcode`\'=\active \def'{{}^\bgroup\prim@s}}

\catcode`@=12

\newcommand{\bq}{\begin{equation}}
\newcommand{\be}{\begin{equation}} 
\newcommand{\fq}{\end{equation}}
\newcommand{\ee}{\end{equation}}
\newcommand{\bqr}{\begin{eqnarray}}
\newcommand{\beqs}{\begin{eqnarray}} 
\newcommand{\fqr}{\end{eqnarray}}
\newcommand{\eeqs}{\end{eqnarray}}

\newcommand{\rf}[1]{(\ref{#1})}

\def\bop#1{\setbox0=\hbox{$#1M$}\mkern1.5mu
	\vbox{\hrule height0pt depth.04\ht0
	\hbox{\vrule width.04\ht0 height.9\ht0 \kern.9\ht0
	\vrule width.04\ht0}\hrule height.04\ht0}\mkern1.5mu}

\begin{document}
\thispagestyle{empty} 

\begin{flushright} 
\begin{tabular}{l} 
ANL-HEP-PR-01-020 \\
hep-th/0103225 \\ 
\end{tabular} 
\end{flushright}  

\vskip .3in 
\begin{center} 

{\Large\bf Cosmological constant in broken maximal sugras} 

\vskip .3in 

{\bf Gordon Chalmers} 
\\[5mm] 
{\em Argonne National Laboratory \\ 
High Energy Physics Division \\ 
9700 South Cass Avenue \\ 
Argonne, IL  60439-4815 } \\  
 
{e-mail: chalmers@pcl9.hep.anl.gov}  

\vskip .5in minus .2in 

{\bf Abstract} 
 
\end{center}
\noindent
We examine the form of the cosmological constant in the loop expansion 
of broken maximally supersymmetric supergravity theories, and after 
embedding, within superstring and M-theory.  Supersymmetry breaking at 
the TeV scale generates values of the cosmological constant that are in 
agreement with current astrophysical data.  The form of perturbative 
quantum effects in the loop expansion is consistent with this parameter 
regime.  

\setcounter{page}{0} 
\newpage 
\setcounter{footnote}{0} 

\section{Introduction} 

Data from type I redshifted supernova indicate a small but non-vanishing value 
of the cosmological constant \cite{Perlmutter:1997hx,Riess,Carroll:2000fy}.  
The cosmic microwave background and its anisotropy support this experimental 
evidence.\footnote{These data are subject to error bars, and we fit to the 
quoted value, even though the cosmological constant is arguably consistent 
with zero experimentally.}  A non-vanishing value of $\Lambda_{c}$ is a 
theoretical challenge to reconcile with supergravity and superstring (M-) 
theory, and in this letter we examine this problem in the context of maximal 
supersymmetry, both spontaneously broken \cite{Scherk:1979ta,Cremmer:1979uq} 
and generally broken.  In the former theories a quartic mass relation found 
by tuning the moduli of the compactification may enforce the leading cancellation 
to all loop orders.  In the latter theories, breaking supersymmetry in a manner in 
which 
there are eight independent supersymmetry breaking scales excludes the quartic 
$\Lambda^4$ and sextet $\Lambda^6$ term in the expansion of the cosmological 
constant as we examine in this work; the remaining terms in the expansion are 
in agreement with data.   

There are two forms of the cosmological constant problem: string scale and 
low scale.  In this work we analyze the issue of breaking supersymmetry and 
solving the cosmological constant problem at the order of a TeV.  

The cosmological constant in maximally supersymmetric gravity theories has 
been examined in detail up to one loop \cite{Cremmer:1979uq,Rohm:1984aq}.  
Spontaneously broken supergravity theories are parameterized by four real 
numbers, characterizing the mass spectrum \cite{Ferrara:1979fu}; we also 
consider more general scenarios, as this susy breaking does not contain 
eight independent scales in four-dimensions in association with the number 
of supercharges.  The former example contains several mass relations such 
that the graded trace up to cubic order of the particle masses 
\cite{Cremmer:1979uq,Ferrara:1979fu}, 
\bqr  
\sum (-)^F ~m_i = \sum (-)^F ~m_i^2 = \sum (-)^F ~m_i^3 = 0 \ , 
\fqr 
must be equal to zero.  
There is a fourth relation \cite{Rohm:1984aq} beyond that in \cite{Cremmer:1979uq}, 
\bqr 
\sum (-)^F ~m_i^4 = 0 \ .  
\fqr 
The cosmological constant is finite to one-loop order and has the form 
\cite{Rohm:1984aq}, 
\bqr 
A = \sum_j (-)^F \left( m_j^4 \ln({m_j^2/\Lambda^2}) - {3\over 2} m_j^4 \right) = 
  \sum_j (-)^F~ m_j^4 \ln{m_j^2} \ . 
\label{oneloop} 
\fqr 
A further relation of the mass spectrum sets this also to zero.  This work will 
derive similar results on the ultraviolet nature of the cosmological constant 
with very mild analyticity requirements imposed on the functional form of the 
cosmological constant in the loop expansion.  The series is given and the form 
is shown to agree remarkably well with astrophysical data.  

Maximal supergravity theories \cite{Cremmer:1978km,deWit:1977fk,deWit:1982ig} 
possess thirty-two conserved components or eight Weyl supercharges in four 
dimensions.  The cosmological constant induced after breaking supersymmetry 
is generically  large in gravitational theories, and of the order $\Lambda^4$ 
with $\Lambda$ the Wilsonian ultra-violet  cutoff.\footnote{A review of the 
cosmological constant in gravitational theories may be found in \cite{Weinberg}}  
However, if we demand that the leading order term in the coupling expansion of 
the constant vanishes upon restoration of any two amounts of the supersymmetry 
($N=1$ supersymmetry without matter contains a $\Lambda^6/m_{pl}^2$ term from 
the D-terms), then the leading unique polynomial describing the cosmological 
constant is, 
\bqr   
{\cal L}=\int d^dx \sqrt{g} ~{1\over m^4_{\rm pl}} \prod_{j=1}^8 \Lambda_j \ ,
\label{leadingorder}
\fqr 
in which $\Lambda_j$ defines the scale of non-conservation of the independent 
supercharges.  The cancellations of the $\Lambda^4$ and $\Lambda^6/ 
m_{\rm pl}^2$ terms are associated with approximate supersymmetry 
restoration at high-energies together with the extraction of eight powers of loop 
momentum in the tensor integrations of the ${\cal N}=8(32)$ theory.  The particle 
masses are taken to be fractions of the supersymmetry breaking scale, $m_j=\alpha_j 
\Lambda$ and the scales as $\Lambda_j=\beta_j \Lambda$.   The two distinct energy 
scales are associated  with : 1) the supersymmetry breaking scales $\Lambda_j$, 2) 
the gravitational coupling constant which is taken to be the Planck mass (with the 
same for the UV cutoff in the field theory).  

The term in \rf{leadingorder} vanishes upon restoration of any two (or one) of the 
conserved supercharges.  Having only seven independent supersymmetry breaking scales 
would allow the function 
\bqr 
(c_1 \Lambda_1\ldots \Lambda_6 + {\rm perms})/m_{pl}^2 
\fqr 
which would not vanish upon taking one of the supersymmetry breaking scales 
to be zero.  This is compatible with the $\Lambda^6/m_{pl}^2$ term in $N=1$  
supergravity. If there are eight independent scales (all of which are present 
in the expansion), then this sextet term would vanish upon restoration of 
any two supersymmetries.  

The numerical value, modulo a suppressed coefficient of order unity, 
agrees very well with the experimental value of the cosmological constant when the 
supersymmetry breaking is at the TeV scale, 
\bqr  
{\Lambda^8\over m_{\rm pl}^4} \sim 2\times 10^{-10} {\rm erg/cm^3}  \ , 
\label{cosmoconstant}
\fqr   
with $\Lambda\sim 3 \times 10^3$ GeV (and $m_{\rm pl}=10^{19}$ GeV).  The constant 
is sensitive 
to the supersymmetry breaking scale, which receives a $2^8$ factor from doubling 
all the scales.  The naive leading order terms,   
\bqr 
\alpha\Lambda^4 \qquad\quad \beta {\Lambda^6\over m_{\rm pl}^2}  \ ,
\fqr 
generically present, are absent in this supersymmetry breaking.  In the 
spontaneously broken ${\cal N}=8$ theory, mass relations enforce these terms to 
vanish.

We consider next the series expansion of the cosmological constant in the 
general maximal supersymmetric theories with multiple supersymmetry breaking 
scales $\Lambda_j$; we do not give the explicit mechanism in this work, however 
explicit supersymmetry breaking terms.  The effective 
action we consider is obtained by integrating modes up to the highest energy 
scale $\Lambda_1<\Lambda_2 \ldots < \Lambda_8$, with no particular range of 
the $\Lambda_j$ (such as holding all $\Lambda_j \ll \Lambda_8$).  Then the physics
of the theory is  governed by $\Lambda_j$ together with, in the field theory, the 
ultra-violet Wilsonian cutoff $\Lambda$, and the masses $m_j$ of the particles (if 
considering the string).  The gravitational coupling constant is $m_{\rm pl}$.  
In the string models, we may simply induce the supersymmetry breakings below the 
string scale.

We analyze the form according to the two sectors : 1) loop diagrams 
containing a propagating gravitational mode with coupling $m_{\rm pl}^{-2}$ 
and 2) loop diagrams containing only the matter degrees of freedom.    
(Theories with spontaneously broken supersymmetries of the Scherk-Schwarz type, 
possessing four scales may, up to the $L^{\rm th}$-loop order 
have canceled quartic and sextet terms if we consider mass relations similar 
to the one in \rf{oneloop}.  This mass relation is the only condition we impose 
on the spontaneously broken expansions, and we assume that a solution exists for 
mass parameters in these breaking scenarios.)    

Inverse powers of the susy breaking scale do not occur in the 
expansions of diagrams containing massive particles, as analyzed below.  
The primitive divergences of multi-loop diagrams are polynomials in $\Lambda$, 
modified by logarithms, and upon taking the particle masses to be proportional 
to $\Lambda$, there are no dimensionless expansion parameters; the  ratios 
$\Lambda_i/\Lambda_j$ within logarithms are of order unity upon taking the 
scales to be the same and do not produce an expansion variable.  At values of 
$\Lambda_j= \beta_j \Lambda$ this enforces the general expansion to be of the 
form, (with the parameters $\alpha_j$ labeling masses in terms of scales) 
\bqr  
{\cal L}_{N=8} = \sum_{n=0}^\infty~ \int d^4x \sqrt{g} ~  c_n(\beta_j,\alpha_j) 
{\Lambda^8\over m^4_{\rm pl}} \Bigl( {\Lambda\over m_{\rm pl}}\Bigr)^{2n}  \ ,
\label{powerseries}
\fqr 
in which supersymmetry breaking is parameterized by eight independent scales.  
Incorporating a particle mass $M$ into the expansion alters the form of the 
$n^{\rm th}$ coefficient into the expansion and generates contributions, 
\bqr  
c_n \Lambda^4 \left({\Lambda^2\over m^2_{\rm pl}}\right)^n \bigl[ \alpha_0^{(n)} 
 {\Lambda^4\over m_{\rm pl}^4} + \alpha_1^{(n)} {\Lambda^4\over m_{\rm pl}^4} 
 \ln^L{\Lambda\over M} + \ldots + 
\beta_0 {\Lambda^3 M\over m_{\rm pl}^4} + \ldots\bigr] + \ldots 
\nonumber 
\fqr 
\bqr 
+ c_n M^4 \bigl[ \beta_0^{(n)} 
 \bigl({M\over m_{\rm pl}}\bigl)^{(2+2n)} \ln^L{\Lambda\over M} + 
 \ldots({\rm log~modifications}) \bigr] + \ldots \ . 
\label{gravterms} 
\fqr 
with $M\sim \Lambda_j$.  The form of the expansion is controlled by the fact 
that every diagram is multiply differentiable in the masses of the internal 
propagators, 
\bqr 
\prod \left\{ \partial\over\partial m_{\sigma(j)} \right\} A_{L~loops}  
\vert_{M=0} \ , 
\fqr 
which exists and has the structure in field theory of 
\bqr 
{\prod m_{\sigma(j)}\over m_{pl}^{\rho_1} \Lambda^{\rho_2} } \ ,  
\label{locality}
\fqr  
with $\Lambda$ the ultra-violet cutoff.  In field theory we take this scale to be of 
order of the $m_{pl}$.  In string theory the form is the same with $\Lambda=m_{pl}$.  
This form in \rf{locality} shows that the loop expansion of the cosmological 
constant is a Taylor series expansion in the supersymmetry breaking scales.  

Next we comment on the loop expansion and the tower of terms in \rf{powerseries} 
for the gravitational sector followed by the matter sector.  The iteration of a 
$L$-loop contribution to $L+1$-loops follows by inserting an additional propagator 
and two graviton three-point vertices, or by higher-point vertices.  The gravitational 
vertices are found by expanding the Einstein-Hilbert term, $m_{\rm pl}^2 
\int d^4x\sqrt{g}~ R$, and we do not list the diagrammatic rules here.  The insertion 
of two three-point vertices introduces four-derivatives into the loop integration 
(the graviton $n$-point vertices contain two derivatives), two propagators, and an 
additional loop integration.  This iteration is 
\bqr 
A_{L+1}= {1\over m_{\rm pl}^4} \int {d^4l\over (2\pi)^4} ~ 
A_{\rm L-1}^{\mu_1\nu_1,\mu_2\nu_2}(l,l+p) \quad \times 
\nonumber
\fqr 
\bqr 
V_{\mu_1\alpha_1\beta_1,\nu_1\gamma_1\delta_1}(l,l+p)  
 V_{\mu_2\alpha_2\beta_2,\nu_2\gamma_2\delta_2}(-l,-l-p) 
 \Delta^{\alpha_1\gamma_1;\alpha_2\gamma_2}(l+p) 
 \Delta^{\beta_1\delta_1;\beta_2\delta_2}(l) \ . 
\label{iterationthreepoint}
\fqr 
and with a four-point vertex is,  
\bqr  
A_{L+1} = {1\over m_{\rm pl}^4} \int {d^4l\over (2\pi)^4} 
A_{\rm L}^{\mu\nu;\alpha\beta}(l,-l)  \Delta^{\mu\nu;\alpha\beta}(l) \ . 
\label{iterationfourpoint} 
\fqr 
The integral in \rf{iterationthreepoint} generates $\Lambda^{4-6+4}=\Lambda^2$ 
additional occurences of the ultra-violet cutoff.  The two graviton three-point 
vertices come with $1/m_{\rm pl}^2$ as the gravitional coupling constant is in 
terms of the Planck mass.  Thus the relative weight between $L$-loops and $L+1$ 
loops differs by a factor of $\Lambda^2/m_{\rm pl}^2$.  The power series in 
\rf{powerseries} reflects the loop expansion in this manner.  Higher point vertices 
do not change the form in \rf{powerseries}, but rather mix different loop orders 
in the coefficients $c_n$, as may be deduced by similar counting.  For example, the 
four-point vertex iteration in \rf{iterationfourpoint} gives the power count 
$\Lambda^{2+4-4}$, with two powers of $\Lambda$ from the derivatives from the 
vertex, four from the loop integration, and minus two from the propagator.  
Higher-point vertices and other particle modes do not change the analysis.  

The pure matter sector, which does not contain explicit occurences of the gravitational 
coupling, may be similarly analyzed.  The expansion of the diagrams in the pure 
matter sector, however, do possess inverse powers of the gravitational coupling upon 
expanding the diagrams at low-energy about an ultra-violet cutoff on the order of 
the Planck mass, as in \rf{locality}.  We consider iterating an $L$-loop scalar 
field  theory graph in $\lambda_1(\Lambda)^{ijk} \phi_i\phi_j\phi_k + \lambda_2 
(\Lambda)^{ijkl} \phi_i\phi_j \phi_k\phi_l$ theory.  Gluing two cubic vertices 
and three propagators iterates a $L$-loop graph to $L+1$ loops, but does not 
raise the degree of primitive divergence.  The logarithmic factors increase by 
one unit, however, in the primitive divergence.  The form of the result is 
\bqr  
c_{L} \bigl( m_i^4 \ln^L m_i + \ldots + m_i^2 \Lambda^2 + \ldots + \Lambda^4 
+ \ldots \bigr) \ , 
\label{matterterms}
\fqr 
and subleading powers in the cutoff, generating terms containing factors of 
$m_{pl}^{-4+k}$.  The indices on the masses represent arbitrary 
combinations.  The 
four-point vertices in the scalar field theory give the same result.  An iteration 
to higher loop order containing a gravitational 
mode with two couplings and a factor $m_{\rm pl}^{-2}$; these terms are 
classified in \rf{gravterms} and are explicitly suppressed by an acceptable power 
of the Planck mass.  The functions \rf{matterterms} to $L$ loop orders are in 
principle computable.

The expansion parameter in the quantum series is $\Lambda/m_{\rm pl}$ and is 
a small dimensionless number, $10^{-15}$.  The series in \rf{powerseries} makes 
sense as an expansion in loops or coupling for this reason, and the leading value 
in \rf{cosmoconstant} is stable under quantum corrections.  

For comparison we discuss the form of the cosmological constant in general 
non-supersymmetric examples (for example, breaking ${\cal N}=1$ supersymmetry).  
The loop expansion contains the series with the two additional terms $\Lambda^4$ 
and $\Lambda^6$, 
\bqr  
{\cal L} = \int d^4x \sqrt{g}~ \Bigl[ \alpha\Lambda^4 +  
 \beta {\Lambda^6\over m^2_{\rm pl}}\Bigr] + \sum_{n=0}^\infty~ \int d^4x \sqrt{g} ~  
 c_n {\Lambda^8\over m^4_{\rm pl}}  \Bigl( {\Lambda\over m_{\rm pl}}\Bigr)^{2n}  \ . 
\fqr 
These two additional terms do not agree with experimental values of the 
cosmological constant without renormalization. 

The loop expansion of the supersymmetrically broken ${\cal N}=8$ supergravity or 
IIB superstring theory naturally produces values of the cosmological constant, 
and a loop expansion, that is in agreement with current astrophysical data if the 
susy scales are of the order of a TeV.  This analysis carries through in the two 
scenarios:  1) susy breaking scales are all independent, in which case \rf{powerseries} 
holds by imposing analytic requirements, and 2) in spontaneous susy breaking in which 
mass relations are required to have \rf{powerseries}.  In the former approach no 
tuning is required.  In both examples, the non-vanishing cosmological constant arises 
from both the gravitational and matter sector. 

M-theory has ${\cal N}=1$, $d=11$ supergravity as its low-energy limit, which describes 
M-theory graviton scattering to high orders in the derivative expansion, and  
contains ${\cal N}=8$ supergravity upon toroidal compactification or dimensional 
reduction.  The symmetries of the maximally supersymmetric theory are 
responsible for the leading order cancellations of terms that spoil 
cosmological predictions of supergravity theories.  Furthermore, the quantum 
corrections do not alter the semi-classical prediction.   

The ${\cal N}=8$ maximally supersymmetric theory, and the IIB superstring (M-) 
theory that contains it as its massless sector, have two properties that have 
been explored recently: 

1) The theory appears finite in perturbation theory according to the modular 
properties of the scattering inherited from S- and U-duality and the AdS/CFT duality 
\cite{Chalmers:2000zg,Chalmers:2000vq,Chalmers:2000ks} (after decoupling of the 
massive modes).  The cancelations resulting in finiteness may occur in a string 
inspired regulator preserving these properties, and are beyond the cancelations 
at two-loops in the expansion of the graviton scattering 
\cite{Bern:1998ug,Bern:2001zc}.  

2) Upon breaking supersymmetry at the TeV scale, the theory produces a cosmological 
constant that agrees with current experimental data and which is stable under 
quantum corrections.  

\noindent
The two properties warrant further phenomenological investigations of both 
IIB superstring (and M-) and ${\cal N}=8$  theories, and in particular to answer 
if the standard model may be accomodated in its low-energy physics.  

\section*{Acknowledgements} 

The work of GC is supported in part by the US Department of Energy, Division of High 
Energy Physics, contract W-31-109-ENG-38.  GC thanks Gia Dvali and Emil Martinec for 
relevant discussions and correspondence.

\end{document}